\documentclass[12pt]{article}
\usepackage{newtxtext,newtxmath}
\usepackage{graphicx}
\usepackage[letterpaper,margin=1in]{geometry}
\usepackage{cite}
\usepackage[colorlinks=true,linkcolor=blue,citecolor=blue,urlcolor=blue]{hyperref}
\usepackage{url}
\usepackage{amsmath,amsfonts,bm}
\usepackage[switch]{lineno}

\renewenvironment{abstract}{\quotation}{\endquotation}
\date{}

\makeatletter
\renewcommand{\fnum@figure}{\textbf{Figure \thefigure}}
\renewcommand{\fnum@table}{\textbf{Table \thetable}}
\renewcommand\@cite[2]{\textsuperscript{\,[#1]}}
\makeatother

\newcommand{\diag}{\operatorname{diag}}
\newcommand{\R}{\mathbf{R}}
\newcommand{\U}{\mathbf{U}}
\newcommand{\A}{\mathbf{A}}
\newcommand{\B}{\mathbf{B}}
\newcommand{\F}{\mathbf{F}}
\newcommand{\X}{\mathbf{X}}
\newcommand{\I}{\mathbf{I}}
\newcommand{\G}{\mathbf{G}}
\newcommand{\Gt}{\widetilde{\mathbf{G}}}
\newcommand{\Lam}{\boldsymbol{\Lambda}}

\def\scititle{Conservation-Based Feedback-Circuit Decomposition for Linear Forced Systems}
\title{\bfseries \boldmath \scititle}
\author{
Ming Cai$^{1*}$\\[0.3cm]
{\small $^1$Department of Earth, Ocean, and Atmospheric Science}\\
{\small Florida State University, Tallahassee, FL 32306, USA}\\[0.2cm]
{\small *Corresponding author: mcai@fsu.edu}
}

\begin{document}
\maketitle

\begin{abstract} \bfseries \boldmath
We present a conservation-based feedback-circuit decomposition specifically for general linear forced systems. In a role parallel to that of eigenvalues and eigenvectors for initial-value problems, the complete set of independent intrinsic circuit gains and their associated forcing-transformation vectors provide a complete analytical representation of both transient and equilibrium forced solutions. The sign of intrinsic circuit gains determines whether successive feedback cycles exhibit monotonic or oscillatory convergence to transformed forcing, while the forcing-transformation vectors determine the structure of transformed forcing. The exact transient and equilibrium solutions are represented analytically through the convergence of the finite-cycle forcing-transformation kernel to the equilibrium forcing-transformation kernel, which is guaranteed regardless of whether the magnitudes of circuit gains exceed one or unstable modes exist in the system. The feedback-circuit decomposition provides a new generic foundational mathematical tool for understanding, predicting, and controlling forced responses in a broad range of coupled linear systems across science and engineering.
\end{abstract}

\section*{Introduction}

Coupled linear systems arise throughout the natural sciences, engineering, and applied mathematics, including climate dynamics, fluid systems, electrical circuits, biological networks, and complex interacting systems. Two fundamental classes of problems commonly emerge in the analysis of such systems: initial-value problems describing the temporal evolution of the system in the absence of external forcing, and forced problems describing the system response to externally imposed perturbations. Although both classes of problems are fundamental to modern science, physically interpretable decomposition frameworks have developed much more extensively for initial-value problems than for forced problems.

For initial-value problems, eigenvalues and eigenvectors provide a general framework for representing the intrinsic evolution of linear systems \cite{strang2006,courant1989,strogatz2015}. These modal decompositions characterize the temporal growth, decay, oscillation, and propagation of linearly coupled disturbances, and have become foundational tools across fluid dynamics, wave theory, quantum mechanics, control theory, atmospheric dynamics, and many other branches of science and engineering. Through eigenmode representations, the evolution of a coupled linear system can be interpreted in terms of independent dynamical modes with distinct temporal and spatial structures. A broad range of related modal and operator-based frameworks have subsequently been developed, including generalized normal modes \cite{farrell1988,trefethen2005}, singular vectors \cite{palmer1998,golub2013}, dynamic mode decomposition \cite{schmid2010}, Koopman operator analysis \cite{rowley2009,mezic2005}, and related spectral representations. Although these methods differ in formulation and application, they share the common objective of identifying dynamically intrinsic structures that provide physically interpretable representations of system evolution.

Forced coupled linear systems are also commonly represented through well-established mathematical frameworks, including inverse operators and functional-analytic formulations \cite{conway1990,kato1976}, Green's functions \cite{economou2006}, propagator, semigroup, and resolvent formulations \cite{engel2000,hille1957}, feedback-control theory \cite{astrom2008,franklin2015}, and related matrix representations. These approaches provide rigorous mathematical representations and exact solution frameworks for externally forced responses and have become central tools across the physical and applied sciences. In Green's function methods, the inverse operator characterizes the equilibrium response to externally imposed perturbations \cite{economou2006}. Influence-matrix formulations provide analogous representations in inverse problems \cite{penrose1955,tarantola2005}. In classical feedback and control theory, externally designed feedback loops are widely used to modify the transient and equilibrium responses of forced systems \cite{astrom2008,franklin2015}. Series expansions of inverse operators, including Neumann-series representations, have also been widely used across mathematics, physics, and engineering to characterize the propagation of forcing through coupled interaction pathways \cite{kato1976,varga1962}. In Gaussian graphical models and related network formulations, such expansions have further been interpreted as walk-sum representations of coupled forced responses \cite{malioutov2006}.

Despite these extensive developments, a general physically interpretable decomposition framework specifically for linear forced systems analogous to eigenmode decompositions for initial-value problems has remained lacking. Here we show that an invertible linear system with nonzero diagonal entries and nonsingular principal submatrices admits an exact analytical decomposition into a complete set of independent intrinsic feedback circuits. This complete set of independent intrinsic feedback circuits analytically describes how externally imposed forcing evolves through forcing-transfer pathways and successive feedback cycles governed by intrinsic circuit gains. The forced response at finite feedback cycles is represented through a finite-cycle forcing-transformation kernel that characterizes progressive forcing transformation and the balancing of transformed forcing as equilibrium is approached. At equilibrium, the convergence of the finite-cycle forcing-transformation kernel through successive feedback cycles gives rise to the equilibrium forcing-transformation kernel representing the equilibrium transformation of external forcing across coupled linear system components. The resulting framework therefore provides an exact analytical representation of coupled linear forced systems in terms of intrinsic forcing-transfer pathways and circuit gains that govern successive balances of transformed forcing at individual system components, analogous to, but fundamentally distinct from, eigenmode representations for initial-value problems.
\vspace{-0.5cm}
\section*{Results}

We consider the forced response for a general linear dynamical system with zero initial condition and time-independent forcing,
\begin{equation}
\frac{d\X}{dt}+\A\X=\F,
\label{eq:system}
\end{equation}
where $\A\in\mathbb{C}^{L\times L}$ is an invertible matrix with nonzero diagonal entries and nonsingular principal submatrices, $\F\in\mathbb{C}^{L}$ is an externally imposed forcing vector, and $\X\in\mathbb{C}^{L}$ is the system state vector. 

\subsection*{Two types of feedback circuits and forcing transfers}

Conventional formulations for linear forced problems solve for the forced response through the fully coupled linear system as a whole, in which forcing is balanced collectively across all system components. Within such formulations, it is generally not possible to explicitly identify transfers of forcing among different system components or subsystems. In the feedback-circuit formulation, the equilibrium balance of the full system is instead represented by separately considering the balancing of forcing at individual system components and within their complementary subsystems. Because the system components remain coupled, the balancing of forcing at one system component necessarily induces transfers of unbalanced forcing to the complementary subsystem and vice versa. By construction, such transferred forcing is not balanced within the receiving subsystem and therefore acts as unbalanced forcing there.

\begin{figure}
\centering
\includegraphics[width=0.78\textwidth]{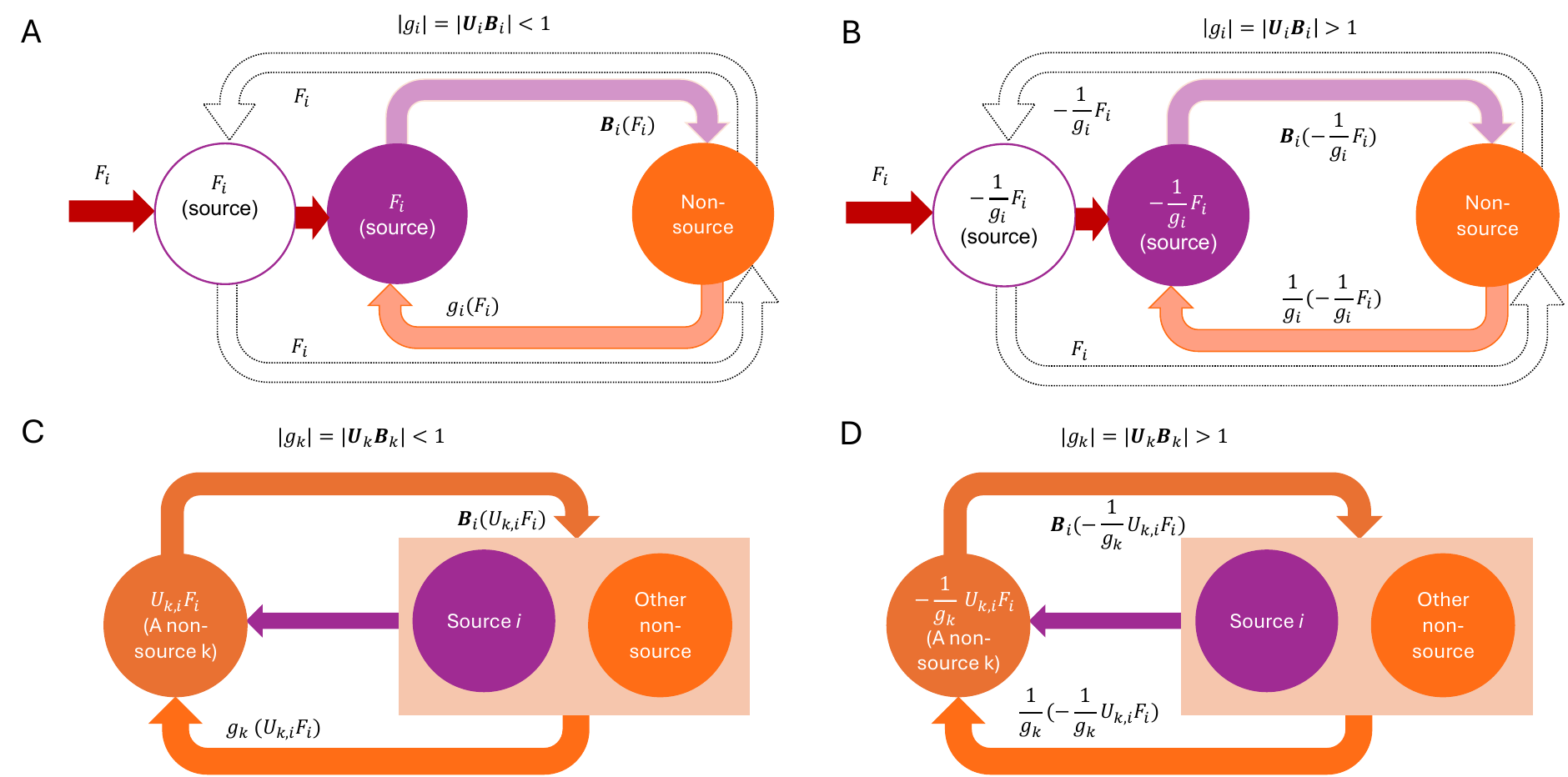}
\caption{\textbf{Illustration of feedback-circuit decomposition of a coupled linear system.}
Panels (\textbf{A}, \textbf{B}) show source-centered feedback circuits, and panels (\textbf{C}, \textbf{D}) show non-source-centered feedback circuits. In (\textbf{A}, \textbf{B}), ``non-source'' denotes all components excluding component $i$; in (\textbf{C}, \textbf{D}), ``other non-source'' denotes all components excluding components $i$ and $k$. The black dashed arrows denote the pre-cycle transfer between the source and non-source components in the source-centered circuit, prior to successive feedback cycles. The colored arrows denote successive transfers of unbalanced forcing during feedback cycles. The intrinsic feedback circuit gain $g_i=\U_i\B_i$ is given by the product of the $i$-th row of $\U$ and the $i$-th column of $\B$. The panels on the left are for the case in which the magnitude of the intrinsic feedback circuit gain is less than one, whereas those on the right are for the case in which the magnitude of the intrinsic feedback circuit gain is greater than one.}
\label{fig:fig_1}
\end{figure}

\begin{figure}
\centering
\includegraphics[width=0.78\textwidth]{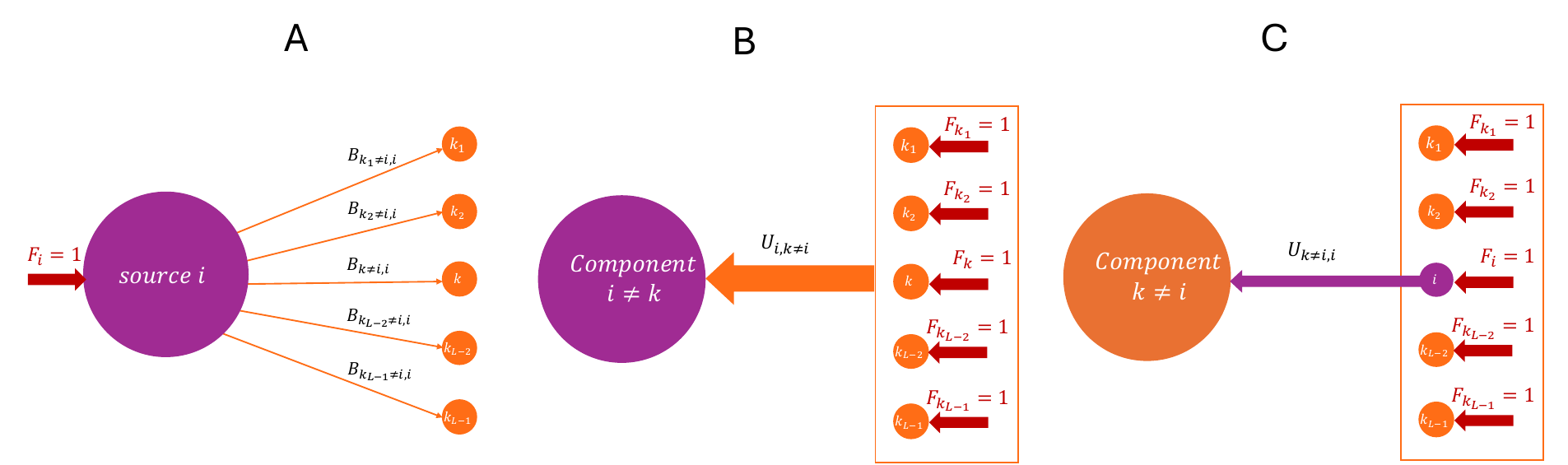}
\caption{\textbf{Illustration of within-circuit and cross-circuit transfers of unbalanced forcing.}
(\textbf{A}) Within-circuit forcing transfer to non-source components ($B_{k\ne i,i}$) associated with a local balance at a source component $i$, where a unit forcing $F_i=1$ is externally imposed. (\textbf{B}) Within-circuit forcing transfer to component $i$ from its complementary subsystem due to the collective non-local balance of unit forcing within the subsystem. This forcing transfer is represented through row-wise ($U_{i,k\ne i}$) transfer factors of matrix $\U$. (\textbf{C}) Cross-circuit forcing transfer to a non-source component $k$ from the source component $i$ specifically, resulting from the collective non-local balance of unit forcing within a subsystem that includes the component $i$ but excludes the non-source component $k$. This forcing transfer is represented through column-wise ($U_{k\ne i,i}$) forcing transfer factors of matrix $\U$.}
\label{fig:fig_2}
\end{figure}

For a given forcing component $F_i$, we refer to the corresponding system component $X_i$ as the source component associated with the externally imposed forcing, and the remaining components as non-source components. Accordingly, as illustrated in Fig.~\ref{fig:fig_1}, the feedback circuit associated with an individual system component serves as a source-centered circuit when considering the externally imposed forcing applied to that component, in which other components are regarded as non-source components. The same circuit serves as a non-source-centered circuit when the source component is one of the components in its complementary subsystem. For easy reference, we refer to the forcing transfer between the centered component and non-centered components as the within-circuit forcing transfer, regardless of whether the circuit serves as a source-centered or non-source-centered circuit. Within-circuit forcing transfers are expressed in two forms. One is for the transfer of unbalanced forcing to each of the non-centered components from the centered component at which forcing is balanced locally, as illustrated in Fig.~\ref{fig:fig_2}A and corresponding to the upper forcing transfer pathway in Fig.~\ref{fig:fig_1}. The other is for the transfer of unbalanced forcing to the centered component from each non-centered component due to their collective non-local balance of the forcing there, as illustrated in Fig.~\ref{fig:fig_2}B and corresponding to the lower forcing transfer pathway in Fig.~\ref{fig:fig_1}. The forcing transfer specifically from the source component to other components is referred to as the cross-circuit forcing transfer, as it only occurs when the circuit serves as a non-source-centered circuit. The cross-circuit forcing transfer represents the unbalanced forcing entering a non-source-centered component from the source component, resulting from the collective non-local balance with the complementary subsystem associated with that non-source-centered circuit. The cross-circuit forcing transfer is illustrated in Fig.~\ref{fig:fig_2}C corresponding to the purple arrow in Figs.~\ref{fig:fig_1}C and~\ref{fig:fig_1}D.

The within-circuit forcing transfer occurring in the upper forcing transfer pathway of Fig.~\ref{fig:fig_1} is characterized by the matrix $\B=(B_{k,i})$, defined by
\begin{equation}
B_{k,i}=-\frac{A_{k,i}}{A_{i,i}},\quad k\ne i,\qquad B_{i,i}=1.
\label{eq:B}
\end{equation}
Matrix $\B$ is otherwise identical to the matrix used in walk-sum formulations of Gaussian graphical models \cite{malioutov2006}, except that the diagonal entries of $\B$ are unity instead of zero. In the feedback-circuit framework, the diagonal elements $B_{i,i}=1$ represent the externally imposed unit forcing, while the off-diagonal elements of the $i$-th column represent the proportional within-circuit transfer of unbalanced forcing to individual non-centered components from the centered component $i$, resulting from the local balance of the unit forcing at the that component.

To obtain the within-circuit forcing transfer occurring in the lower pathway of Fig.~\ref{fig:fig_1} according to Fig.~\ref{fig:fig_2}B, we first construct the matrix $\U$ from the inverse of a reduced matrix obtained by deleting the $i$-th row and $i$-th column of $\A$, denoted by $\left(P^{(i)}_{k',l'}\right)$. The matrix $\left(P^{(i)}_{k',l'}\right)$ is then embedded back into the original $L\times L$ system through the matrix $\left(\widehat{P}^{(i)}_{k,l}\right)$, defined by
\begin{equation}
\widehat{P}^{(i)}_{k,l}=P^{(i)}_{k',l'},\quad
k=\begin{cases}
k', & k'<i,\\
k'+1, & k'\ge i,
\end{cases}
\qquad
l=\begin{cases}
l', & l'<i,\\
l'+1, & l'\ge i.
\end{cases}
\label{eq:Pembed}
\end{equation}
The $i$-th row of the transfer matrix $\U=(U_{i,l})$ is defined by
\begin{equation}
U_{i,l\ne i}=-\sum_{k\ne i} A_{i,k}\widehat{P}^{(i)}_{k,l\ne i},\qquad U_{i,i}=1.
\label{eq:U}
\end{equation}
Repeating this construction for all $i=1,2,\ldots,L$ yields the full transfer matrix $\U$. The $i$-th row of $\U$ corresponds to the proportional within-circuit transfer of unbalanced forcing to component $i$ from its complementary subsystem due to the collective non-local balance of forcing within the subsystem, as illustrated in Fig.~\ref{fig:fig_2}B. According to Fig.~\ref{fig:fig_2}C, the cross-circuit forcing transfer shares an otherwise identical structure to that for the within-circuit transfer illustrated in Fig.~\ref{fig:fig_2}B except the source component is specifically within the complementary subsystem associated with a non-source component. Therefore, the $i$-th column of $\U$ corresponds to the proportional cross-circuit transfer of unbalanced forcing to a non-source component from the component $i$ specifically after its collective non-local balance of forcing within a subsystem that includes the component $i$ but excludes the non-source component under consideration. 

\subsection*{Forcing transfer pathways and feedback gains}

As illustrated in Fig.~\ref{fig:fig_1}, feedback circuits for a coupled linear system share a common structure, consisting of an upper and a lower forcing-transfer pathway, regardless of the type of circuits or the magnitude of feedback gains. It suffices to consider source-centered circuits, which have an explicit pre-cycle forcing-transfer pathway, to illustrate the structure and role of these pathways. The same structure extends to non-source-centered circuits, which are driven by internally transferred forcing and involve an implicit pre-cycle forcing-transfer pathway.

The source-centered circuit is associated with a system component $i$ at which an external forcing $F_i$ is applied while all other components are unforced. The two within-circuit forcing transfers, one through the upper pathway and the other through the lower pathway (Fig.~\ref{fig:fig_1}),  collectively determine the intrinsic feedback circuit gain $g_i$,
\begin{equation}
g_i=\sum_{k\ne i} B_{k,i}U_{i,k},
\label{eq:g}
\end{equation}

The case that $|g_i |<1$ corresponds to a regular circuit as the magnitude of the unbalanced forcing returned to the source component is less than that of the original forcing. In this case, the circuit gain can serve as the loop gain as the geometric series associated with a gain with magnitude less than unity is guaranteed to converge. The case that $|g_i |>1$ corresponds to an over-compensating circuit as the magnitude of the unbalanced forcing returned to the source component exceeds that of the original forcing. To preserve the magnitude of the within-circuit forcing transfer while the corresponding geometric series is still convergent, we introduce loop gain $\widetilde{g_i}$, and pre-cycle forcing-transfer factor $R_i$ for an over-compensating circuit, given by

\begin{equation}
\widetilde{g}_i=
\begin{cases}
g_i, & |g_i|<1,\\
1/g_i, & |g_i|>1,
\end{cases}
\label{eq:gtilde}
\end{equation}

\begin{equation}
R_i=
\begin{cases}
1, & |g_i|<1,\\
-1/g_i, & |g_i|>1.
\end{cases}
\label{eq:R}
\end{equation}

The expressions in the top row in Eqs.~\ref{eq:gtilde} and~\ref{eq:R} are for a regular (non-over-compensating) circuit and the bottom row for an over-compensating circuit. Therefore, by introducing a loop gain that is the inverse of the circuit gain and a pre-cycle forcing transfer, the full magnitude of the within-circuit forcing transfer associated with an over-compensating circuit remains entirely within the circuit. Only a portion of the original within-circuit transfer determined by the intrinsic circuit gain remains in the circuit through each successive feedback cycle, which is determined by the loop gain. The remaining portion still takes place within the circuit but at the pre-cycle only. It will be proved shortly, the balance of forcing returned to component $i$ through the within-circuit transfer determined by the loop gain takes place within each feedback cycle, whereas that resulting from the pre-cycle within-circuit transfer takes place before the feedback starts. 

\subsection*{Source-centered circuit}

Let $X_i^{(N)}$ denote the response of component $i$ to the forcing entering the source-centered feedback loop after $N$ feedback cycles, including the within-circuit forcing transfer through the pre-cycle pathway, where $N=0$ corresponds to the response to the forcing entering the source component in the pre-cycle pathway. During the first feedback cycle, the forcing balanced locally at the source component is $R_iF_i$, while the unbalanced forcing returned to the source component is $\widetilde{g}_iR_iF_i$. The forcing balanced at the non-source components during the pre-cycle and first cycle are 
$(1-R_i)F_i$ and $(1-\widetilde{g}_i)R_iF_i$, respectively. Therefore, the
total forcing balanced at the non-source components after the first cycle is
\begin{equation}
(1-R_i)F_i+(1-\widetilde{g}_i)R_iF_i.
\label{eq:cyclebalance}
\end{equation}
Repeating the feedback cycle successively gives three related geometric series. The unbalanced forcing entering the source component at successive cycles is
\begin{equation}
R_iF_i,\quad \widetilde{g}_iR_iF_i,\quad \widetilde{g}_i^2R_iF_i,\quad \widetilde{g}_i^3R_iF_i,\ldots.
\label{eq:entryseries}
\end{equation}
Thus, the total unbalanced forcing entering the source component after $N$ cycles is
\begin{equation}
\frac{1-\widetilde{g}_i^N}{1-\widetilde{g}_i}R_iF_i.
\label{eq:source_entryN}
\end{equation}
The corresponding local balance of the forcing at the source component after $N$ feedback cycles is
\begin{equation}
A_{i,i}X_i^{(N)}=\left(1+\widetilde{g}_i+\widetilde{g}_i^2+\cdots+\widetilde{g}_i^{N-1}\right)R_iF_i
=\frac{1-\widetilde{g}_i^N}{1-\widetilde{g}_i}R_iF_i.
\label{eq:source_responseN}
\end{equation}
The forcing balanced at the non-source components at successive cycles is
\begin{equation}
(1-R_i)F_i+(1-\widetilde{g}_i)R_iF_i,\quad 
\widetilde{g}_i(1-\widetilde{g}_i)R_iF_i,\quad 
\widetilde{g}_i^2(1-\widetilde{g}_i)R_iF_i,\ldots.
\label{eq:nonsource_series}
\end{equation}
Its cumulative balance after $N$ cycles is
\begin{equation}
(1-R_i)F_i+(1-\widetilde{g}_i^N)R_iF_i.
\label{eq:nonsource_cum}
\end{equation}
Because $|\widetilde{g}_i|<1$, the returned unbalanced forcing $\widetilde{g}_i^NR_iF_i$ approaches zero as $N\to\infty$. Consequently, the geometric series all converge to equilibrium solutions. The total unbalanced forcing entering the source component at equilibrium is
\begin{equation}
\sum_{n=0}^{\infty}\widetilde{g}_i^nR_iF_i=\frac{R_iF_i}{1-\widetilde{g}_i}.
\label{eq:source_eq_entry}
\end{equation}
Accordingly, the equilibrium forcing balanced locally at the source component is
\begin{equation}
A_{i,i}X_i=\frac{R_iF_i}{1-\widetilde{g}_i},
\label{eq:source_eq}
\end{equation}
where $X_i=\lim_{N\to\infty}X_i^{(N)}$. The total forcing balanced at the non-source components through successive feedback cycles, including the pre-cycle contribution, is
\begin{equation}
(1-R_i)F_i+\sum_{n=0}^{\infty}(1-\widetilde{g}_i)\widetilde{g}_i^nR_iF_i
=(1-R_i)F_i+R_iF_i=F_i.
\label{eq:nonsource_eq}
\end{equation}
Therefore, at equilibrium, the progressively transformed forcing through successive feedback cycles is completely balanced locally at the source component, while the externally imposed forcing is completely balanced at the non-source components within the source-centered circuit. Because the balance of the external forcing at the non-source components within the source-centered circuit is in the form of forcing output, it cannot be expressed in terms of the system response. Therefore, we need to consider non-source-centered circuits to determine their system response to the externally imposed forcing through its transfer to non-source components.

\subsection*{Non-source-centered circuit}

Non-source components acquire a forcing from the source component through the cross-circuit forcing transfer, as illustrated by the purple arrow in Figs, 1C and 1D, as well as Fig. 2C. Because the cross-circuit forcing transfer is from the transformed forcing that is already fully balanced at the source component within the source-centered circuit, we only need to consider how the cross-circuit transfer of the forcing from the source component is balanced at non-source components through their own circuits.

Starting from this internally transferred forcing $U_{k,i} F_i$ from the source component to a non-source-center circuit, we next derive the corresponding finite-cycle and equilibrium transformed forcing within that individual non-source-centered circuit. Let $X_k^{(N)}$ denote the response of component $k$ after $N$ feedback cycles to the forcing entering the non-source-centered feedback loop, including the pre-cycle transfer pathway, where $N=0$ corresponds to the transformation of the internally transferred forcing $U_{k,i}F_i$ into $R_kU_{k,i}F_i$ through the pre-cycle pathway. Repeating the feedback cycles successively gives the geometric series of unbalanced forcing entering the component $k$,
\begin{equation}
R_kU_{k,i}F_i,\quad \widetilde{g}_kR_kU_{k,i}F_i,\quad 
\widetilde{g}_k^2R_kU_{k,i}F_i,\quad 
\widetilde{g}_k^3R_kU_{k,i}F_i,\ldots.
\label{eq:non_source_series}
\end{equation}
Thus, the total unbalanced forcing entering the component $k$ after $N$ cycles is
\begin{equation}
\frac{1-\widetilde{g}_k^N}{1-\widetilde{g}_k}R_kU_{k,i}F_i.
\label{eq:non_source_entryN}
\end{equation}
The corresponding local balance of the forcing at the component $k$ is
\begin{equation}
A_{k,k}X_k^{(N)}
=\left(1+\widetilde{g}_k+\widetilde{g}_k^2+\cdots+\widetilde{g}_k^{N-1}\right)R_kU_{k,i}F_i
=\frac{1-\widetilde{g}_k^N}{1-\widetilde{g}_k}R_kU_{k,i}F_i.
\label{eq:non_source_responseN}
\end{equation}
Because $|\widetilde{g}_k|<1$, the term $\widetilde{g}_k^NR_kU_{k,i}F_i$, representing the unbalanced forcing entering component $k$ after $N$ cycles, vanishes as $N\to\infty$. At equilibrium, the transformed forcing associated with the initially internally transferred forcing $U_{k,i}F_i$ is balanced locally at component $k$,
\begin{equation}
A_{k,k}X_k=\frac{R_kU_{k,i}F_i}{1-\widetilde{g}_k},
\label{eq:non_source_eq}
\end{equation}
where $X_k=\lim_{N\to\infty}X_k^{(N)}$.

Through all these non-source-centered circuits for $k\ne i$ (Figs.~\ref{fig:fig_1}C and \ref{fig:fig_1}D), together with the source-centered circuit (Figs.~\ref{fig:fig_1}A and \ref{fig:fig_1}B), we fully account for the balancing of the externally imposed forcing at the source component through successive forcing transfers, returns, and instantaneous local balances throughout the entire coupled linear system.

\subsection*{Finite-cycle and equilibrium forcing-transform kernels}

Repeating the source-centered and non-source-centered feedback-circuit constructions for all $i=1,2,\ldots,L$ yields the finite-cycle forcing-transformation kernel $\Lam^{(N)}$, given by
\begin{equation}
\Lam^{(N)}=\R(\I-\Gt^N)(\I-\Gt)^{-1}\U.
\label{eq:LambdaN}
\end{equation}
In Eq.~\ref{eq:LambdaN}, $\U$ is the forcing-transfer matrix given by Eq.~\ref{eq:U},whose rows represent within-circuit forcing transfer through the lower pathway to the centered component and columns represent cross-circuit forcing transfer from a source component to other components of the corresponding subsystem, $\I$ denotes the identity matrix, $\R$ is the pre-cycle forcing-transfer matrix, defined by
\begin{equation}
\R=\diag(R_1,\ldots,R_L),
\label{eq:Rmat}
\end{equation}
and $\Gt$ is the feedback loop-gain matrix, defined by
\begin{equation}
\Gt=\diag(\widetilde{g}_1,\ldots,\widetilde{g}_L).
\label{eq:Gtmat}
\end{equation}
Both matrices $\Gt$ and $\R$ are determined from the intrinsic circuit gain matrix $\G=\diag(g_1,\ldots,g_L)$, given by Eqs.~\ref{eq:gtilde} and \ref{eq:R}, respectively. According to Eqs.~\ref{eq:source_responseN} and \ref{eq:non_source_responseN}, the finite-cycle system response, $\X^{(N)}$, to the imposed forcing $\F$ can then be expressed as
\begin{equation}
\diag(\A)\X^{(N)}=\Lam^{(N)}\F.
\label{eq:responseN}
\end{equation}
Because $|\widetilde{g}_i|<1$ for all feedback circuits, $\Gt^N\to0$ as $N\to\infty$. Consequently, the finite-cycle forcing-transformation kernel converges to the equilibrium forcing-transformation kernel,
\begin{equation}
\Lam=\R(\I-\Gt)^{-1}\U,
\label{eq:Lambda}
\end{equation}
and the equilibrium forced solution becomes
\begin{equation}
\diag(\A)\X=\Lam\F,
\label{eq:response_eq}
\end{equation}
where $\X=\lim_{N\to\infty}\X^{(N)}$. Substituting the conventional equilibrium forced solution $\X=\A^{-1}\F$ into Eq.~\ref{eq:response_eq}, we obtain
\begin{equation}
\Lam=\diag(\A)\A^{-1}.
\label{eq:Lambda_direct}
\end{equation}
Equation~\ref{eq:Lambda_direct} provides a numerical formula for calculating the forcing-transformation kernel $\Lam$ directly from $\A$. Using the resultant $\Lam$ in Eq.~\ref{eq:Lambda}, one can numerically recover the pre-cycle forcing-transfer matrix $\R$, the forcing-transfer matrix $\U$, the feedback loop-gain matrix $\Gt$, and the intrinsic circuit gain matrix $\G$, without explicitly inverting reduced subsystems of $\A$.

\subsection*{Illustration of feedback-circuit decomposition and comparison with eigenmodes}

\begin{figure}
\centering
\includegraphics[width=0.78\textwidth]{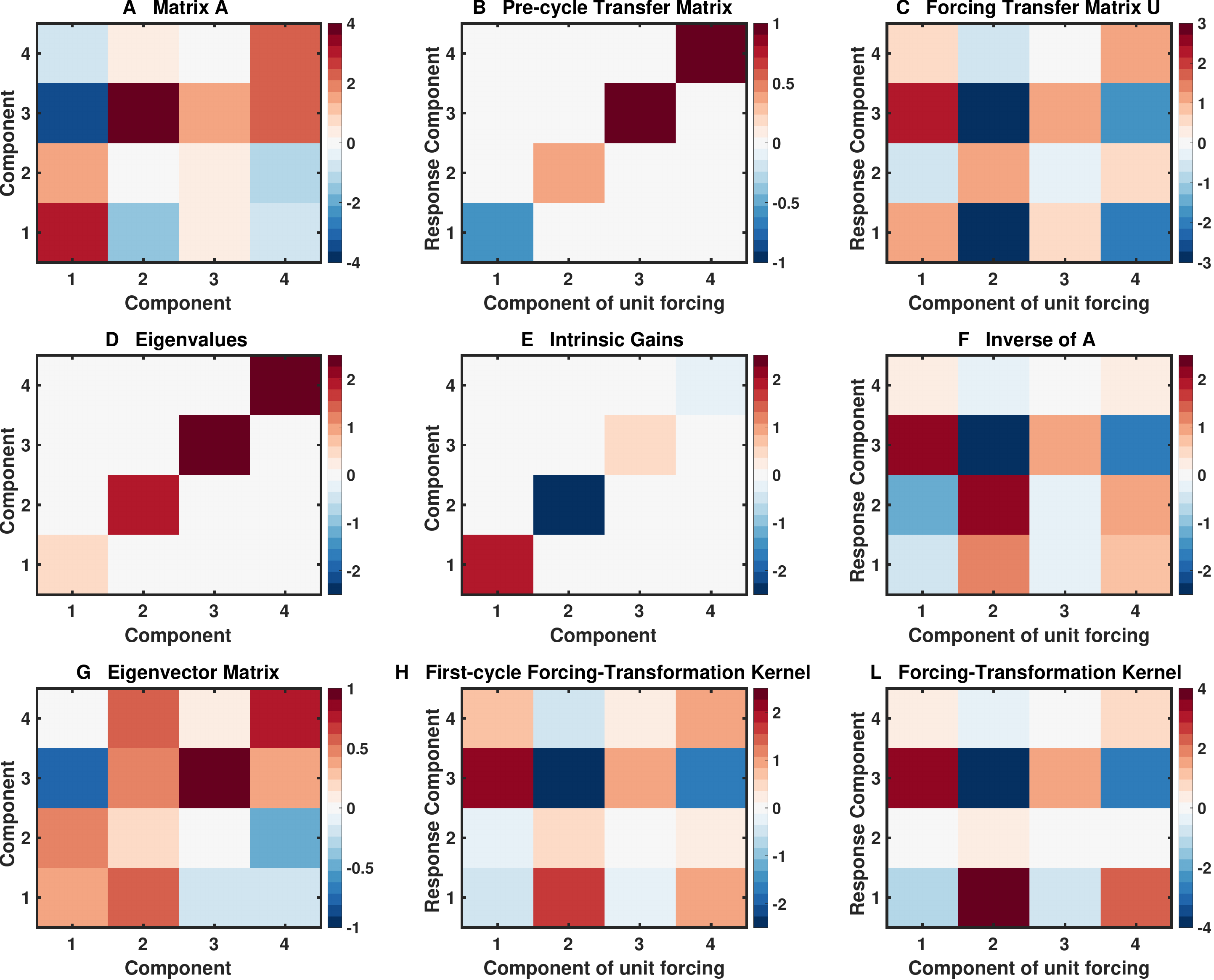}
\caption{\textbf{Example 1 of feedback-circuit decomposition for a stable randomly generated $4\times4$ linear forced system.}
(\textbf{A}) System matrix $\A$. (\textbf{B}) Pre-cycle transfer matrix $\R$. (\textbf{C})  Forcing-transfer matrix $\U$. (\textbf{D}) Eigenvalue matrix. (\textbf{E}) Intrinsic circuit gain matrix. (\textbf{F}) Inverse matrix $\A^{-1}$. (\textbf{G}) Eigenvector matrix. (\textbf{H}) First-cycle forcing-transformation kernel. (\textbf{L}) Equilibrium forcing-transformation kernel. It is noted here that eigenvalues are displayed using the same component ordering as the eigenvectors and circuit quantities for convenience, but this ordering in both coordinates in panel (\textbf{D}), as well as in the horizontal coordinate of panel (\textbf{G}), bears no physical meaning.}
\label{fig:fig3}
\end{figure}

We first illustrate the feedback-circuit decomposition and perform a side-by-side comparison with the eigenmode decomposition. To make this illustration as general as possible, we deliberately choose randomly generated stable and unstable $4\times4$ linear systems that admits intrinsic circuit gains spanning all four representative regimes, namely, gains less than $-1$, between $-1$ and $0$, between $0$ and $1$, and greater than $1$. The illustration of the feedback-circuit decomposition for the stable case is shown in Fig.~\ref{fig:fig3}, while the corresponding convergence of the transient transformed forcing through successive feedback cycles, together with comparisons against direct numerical solutions, is shown in Figs.~S1--S2. The illustration of the feedback-circuit decomposition for the unstable case is shown in Fig.~S3, while the corresponding convergence of the transient transformed forcing through successive feedback cycles is shown in Fig.~4, and direct numerical solutions are given in Fig.~5.

The eigenvalues of the system matrix $\A$ shown in Fig.~\ref{fig:fig3}A are all positive (Fig.~\ref{fig:fig3}D), indicating that the system is dynamically stable. Each eigenvalue corresponds to a global mode of free evolution, while the individual elements of the associated eigenvectors are component specific (Fig.~\ref{fig:fig3}G). In contrast to the eigenmode decomposition, both the intrinsic circuit gains and the forcing-transfer matrices are component specific. The intrinsic circuit gains associated with components 1 through 4 are 1.86, $-2.74$, 0.38, and $-0.18$, respectively (Fig.~\ref{fig:fig3}E). Accordingly, components 3 and 4 exhibit direct positive pre-cycle transfer pathways, with unit pre-cycle transfer factors (Fig.~\ref{fig:fig3}B). By contrast, component 1 has a negative pre-cycle transfer factor because its intrinsic gain exceeds unity, whereas component 2 has a positive but less-than-unity pre-cycle transfer factor because its intrinsic gain is less than $-1$. Because both feedback circuits associated with components 1 and 2 are over-compensating circuits, the corresponding pre-cycle transfer factors have magnitudes less than unity, implying a reduction of the imposed forcing through the pre-cycle forcing transfer.

The matrix $\U$ shown in Fig.~\ref{fig:fig3}C further determines the spatial structure of the forcing transfer among system components during the first feedback cycle. The corresponding first-cycle forcing-transformation kernel shown in Fig.~\ref{fig:fig3}H is obtained from the combined effects of the pre-cycle transfer matrix and the forcing-transfer matrix $\U$, whose mathematical form is $\R$$\U$. Convergence from the first-cycle forcing-transformation kernel (Fig.~\ref{fig:fig3}H) to the equilibrium forcing-transformation kernel (Fig.~\ref{fig:fig3}L) through successive feedback cycles is governed by the corresponding loop gains, which are identical to the circuit gains when their amplitudes are less than unity, and equal to the inverses of the circuit gains when their amplitudes exceed unity. Because the circuit gains for components 1 and 3 are positive, the values in rows 1 and 3 of the forcing-transformation kernel are greater than their counterparts in the first-cycle forcing-transformation kernel, corresponding to monotonic convergence with increasing magnitudes toward transformed forcing in equilibrium (Fig.~S1). By contrast, the negative circuit gains for components 2 and 4 lead to reduced magnitudes in rows 2 and 4 of the forcing-transformation kernel relative to their counterparts in the first-cycle forcing-transformation kernel, corresponding to oscillatory convergence with decreasing magnitudes toward reduced forcing in equilibrium. Other than the magnitude differences, the first-cycle and equilibrium forcing-transformation kernels exhibit highly similar spatial structures, including their sign patterns. The retention of the equilibrium forcing structure is a manifestation of the geometric-series convergence associated with the loop gains, whose magnitudes are always less than unity. The retention of the equilibrium forcing structure by the first-cycle forcing-transformation kernel suggests that the rows of the first-cycle forcing-transformation kernel serve a role analogous to eigenvectors, but for forced problems. Correspondingly, the circuit gains serve a role analogous to eigenvalues by both determining the characteristics of the pre-cycle forcing transfer that contribute to the first-cycle forcing-transformation kernel and governing the convergence rate, as well as whether the magnitude of the transformed forcing at the first cycle is further amplified through monotonic convergence toward equilibrium or reduced through oscillatory convergence toward equilibrium. Fig. S2 shows that the time-integration solutions for the stable case converge to the same equilibrium solutions represented by the feedback-circuit decomposition, although their temporal evolutions are drastically different, governed by the projections of the imposed forcing onto the eigenmodes of the dynamical system.

\begin{figure*}[t]
\centering
\includegraphics[width=\textwidth]{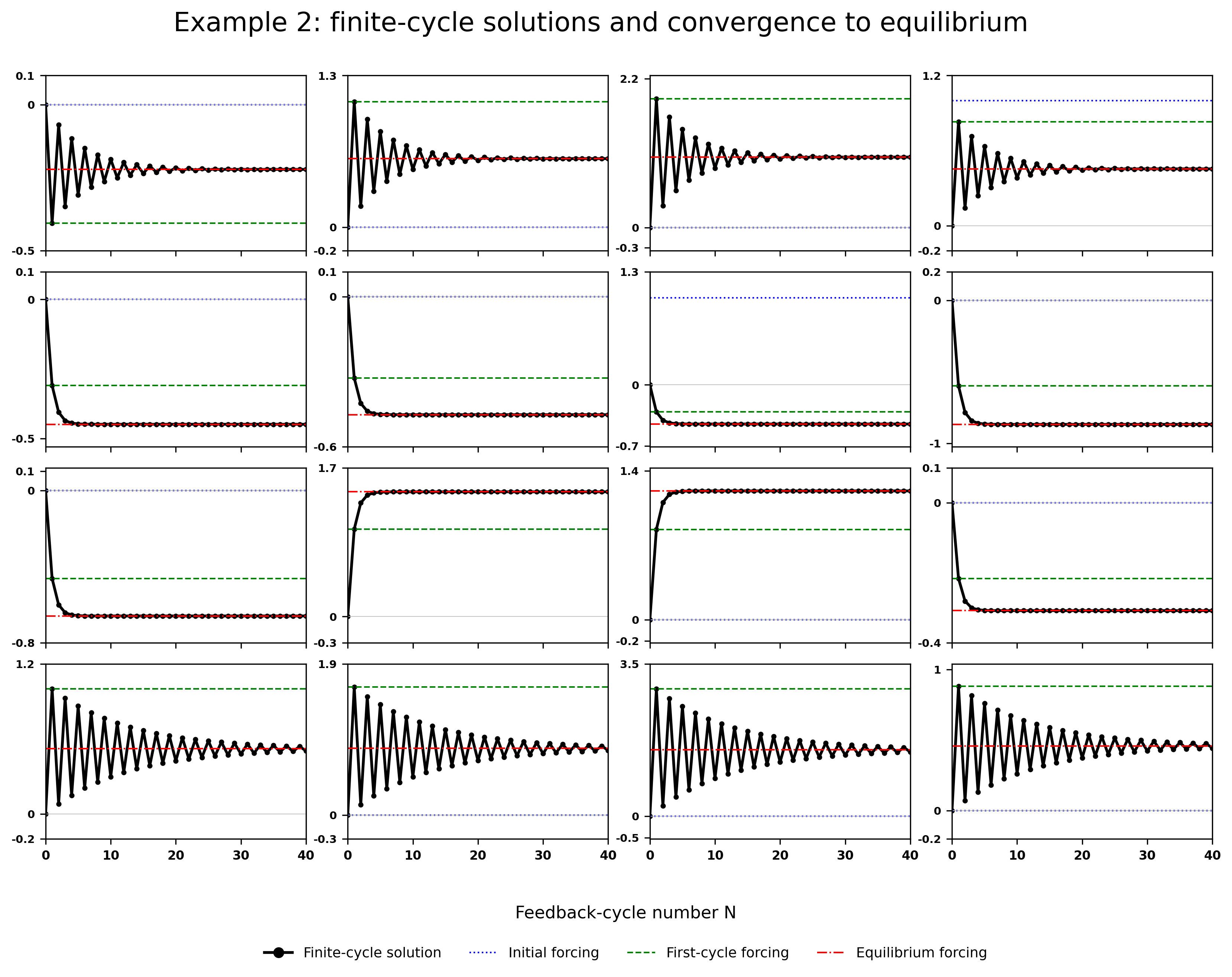}
\caption{
\textbf{Finite-cycle solutions and convergence to equilibrium for Example 2} (an unstable $4\times4$ coupled linear system). Columns represent system components 1--4 from left to right, with the initial input forcing indicated by blue dotted lines. Rows represent the finite-cycle analytical solutions for the transformed forcing (black curves, given by $\boldsymbol{\Lambda^{(N)}F}$ at system components 1--4 from bottom to top in response to the initial forcing imposed at the corresponding column components. The green and red dotted lines denote the analytical first-cycle and equilibrium solutions for the transformed forcing, $\boldsymbol{\Lambda^{(1)}F}$ and $\boldsymbol{\Lambda F}$, respectively. The abscissa represents feedback-cycle number, and the ordinate represents forcing values.
}
\label{fig:example2_cycle}
\end{figure*}

\begin{figure*}[t]
\centering
\includegraphics[width=\textwidth]{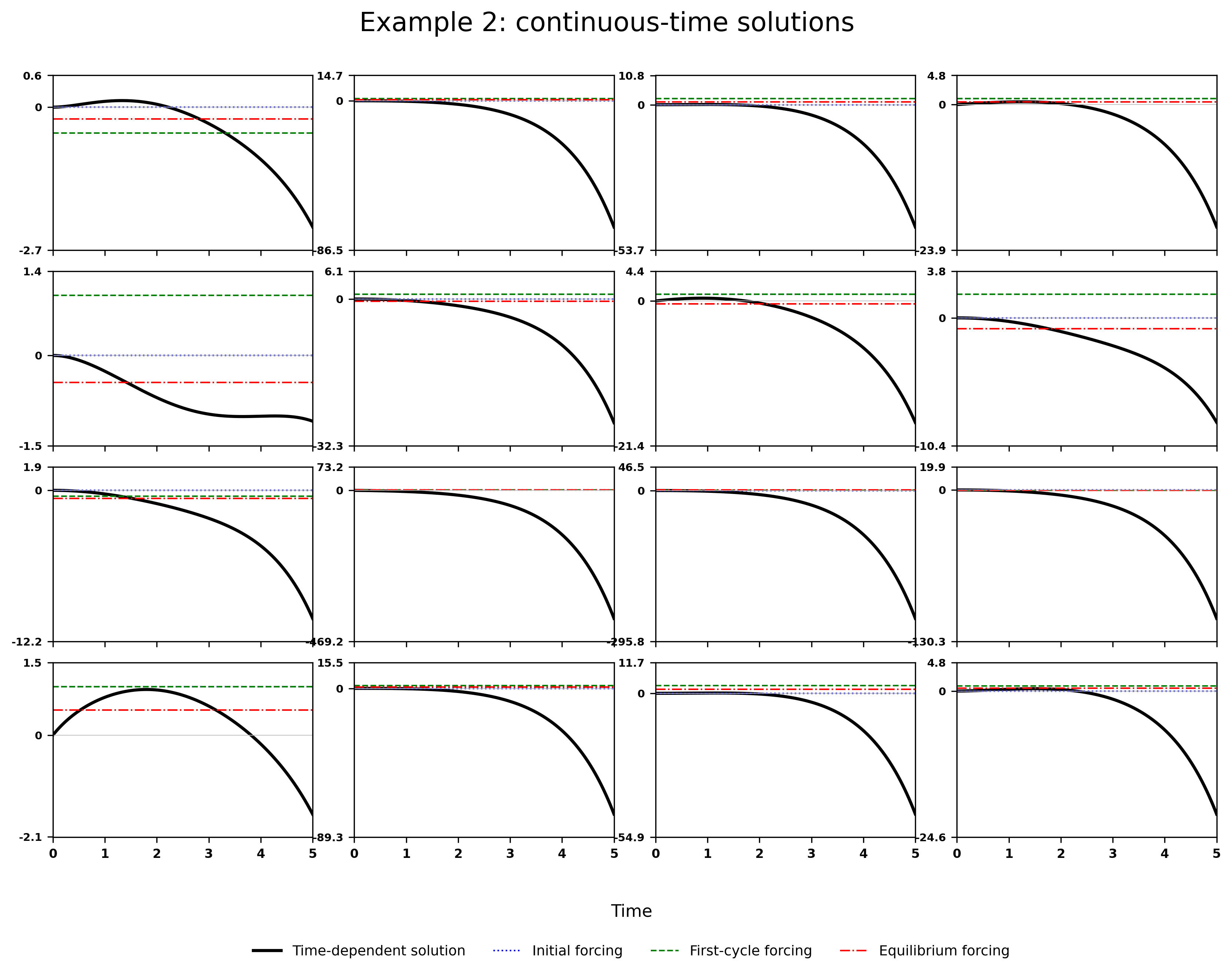}
\caption{\textbf
{Continuous-time solutions and convergence to equilibrium for Example 2} (an unstable $4\times4$ coupled linear system). All information is identical to that in Fig.~\ref{fig:example2_cycle}, except the time-dependent solutions are obtained numerically with the abscissa representing time. Because the numerical solutions do not directly yield the transformed forcing, we deliberately plot $\mathrm{diag}(A)X(t)$ instead of $X(t)$ for a direct comparison with the analytical solutions shown in Fig.~\ref{fig:example2_cycle}.
}

\label{fig:example2_time}
\end{figure*}

The illustration of the feedback-circuit decomposition for the unstable case is shown in Fig. S3. In the main text, we focus primarily on the temporal evolution associated with the unstable example because it provides a particularly clear illustration of the distinction between feedback-cycle evolution and conventional time-dependent dynamical evolution. To illustrate the temporal evolution associated with the unstable example, we design experiments by imposing unit forcing individually at system components 1–4. Columns represent system components 1–4 from left to right, with the imposed unit forcing indicated by the blue dotted lines in the diagonal panels of Figs.~\ref{fig:example2_cycle} and ~\ref{fig:example2_time}. The projections of these imposed forcings onto the unstable eigenmode are approximately -0.03, -1.38, -0.87, and -0.38, respectively. Fig.~\ref{fig:example2_cycle} shows that, despite the existence of an unstable eigenmode, the evolution of transformed forcing at individual system components through successive feedback cycles of the corresponding feedback circuits follows the finite-cycle analytical solutions predicted by the conservation-based feedback-circuit decomposition and converges toward the equilibrium transformed forcing. The spatial structure of the transformed forcing is determined by the combined effects of the pre-cycle forcing transfer and the intrinsic within-cycle forcing transfers, while the temporal evolution is governed by the loop gains associated with the corresponding intrinsic circuit gains. The resulting equilibrium system response determined from the equilibrium transformed forcing is exactly the same as that obtained from the inverse operator.

For example, for the forcing imposed at component 2 (second column in Fig.~\ref{fig:example2_cycle}), the spatial structure of the forcing evolution across the four system components is already established during the first feedback cycle through the combined effects of the pre-cycle forcing transfer and the intrinsic circuit forcing transfer shown in Figs. S3B–C. The corresponding first-cycle forcing-transformation structure shown in Fig. S3H largely determines the spatial structure of the transformed forcing evolution after the first cycle for forcing imposed at component 2, as illustrated by the green dotted lines in the individual panels of the second column in Fig.~\ref{fig:example2_cycle}. The subsequent temporal evolutions of the transformed forcing at each component is governed exactly by the loop gain associated with the corresponding intrinsic circuit gain shown in Fig. S3E. Specifically, the negative loop gains associated with components 1 and 4 produce oscillatory convergence with reducing magnitudes in the first and fourth rows of the second column in Fig.~\ref{fig:example2_cycle} across successive feedback cycles, whereas the positive loop gains associated with components 2 and 3 produce monotonic convergence with increasing magnitude in the second and third rows. Consequently, the convergence of the transformed forcing evolution toward equilibrium is explicitly predictable from the intrinsic transfer pathways and loop gains embedded within the corresponding feedback circuits. The first-cycle forcing-transformation kernel establishes the spatial structure of the transformed forcing entering individual circuits, while subsequent feedback cycles modify their magnitudes according to the loop gains.

In contrast, Fig.~\ref{fig:example2_time} shows that the corresponding continuous-time solutions obtained through direct numerical time integration exhibit diverging behavior because the time-integration evolution is governed by the unstable eigenmode of the dynamical system. Because the projections of the unit forcings imposed at the individual components onto the unstable eigenmode are all negative, the time-integration solutions all diverge toward negative infinity. The speed at which the time-integration solutions diverge away from the equilibrium solutions is largely determined by the projections of the imposed unit forcings onto the unstable eigenmode. In particular, the forcing imposed at component 2 has the largest projection onto the unstable eigenmode, resulting in the fastest diverging among the four columns in Fig.~\ref{fig:example2_time}. The forcing imposed at component 1 has the weakest projection onto the unstable eigenmode, allowing the corresponding time-integration solutions to initially evolve toward the first-cycle transformed forcing structure indicated by the green dotted lines. Nevertheless, the unstable eigenmode eventually dominates the temporal evolution, preventing the time-integration solutions from converging toward the equilibrium solutions. As for the stable case (Fig. S1 versus Fig. S2), the results of Fig.~\ref{fig:example2_cycle} versus Fig.~\ref{fig:example2_time} further demonstrate, but in a more dramatic way, that the temporal evolution predicted by the conservation-based feedback-circuit decomposition is fundamentally distinct from conventional time-dependent dynamical evolution. The former is governed by intrinsic forcing-transfer pathways and loop gains associated with the feedback circuits and always converges to equilibrium, whereas the latter is governed by the eigenmodes of the dynamical system and converges to equilibrium only when the projection of the imposed forcing onto unstable eigenmodes is exactly zero.

We next illustrate the feedback-circuit decomposition using a real-world application from climate science by considering the Planck feedback matrix, which was first proposed by Lu and Cai \cite{lu2009} and Cai and Lu \cite{cai2009} for diagnosing temperature responses to climate forcings. Each column of the Planck feedback matrix represents the vertical profile of thermal cooling rate, in units of W~m$^{-2}$~K$^{-1}$, induced by a 1~K warming in the corresponding layer (Fig.~\ref{fig:fig6}A; see the Supplemental Material for data and numerical details). The diagonal elements of the Planck feedback matrix are positive, indicating the positive thermal emission perturbations resulting from their 1~K warming, whereas all off-diagonal elements are negative, indicating heating perturbations at other layers due to their absorption of emission from the diagonal layer of the corresponding column. All eigenvalues of the Planck feedback matrix are positive (Fig.~\ref{fig:fig6}C), indicating that the climate system is dynamically stable under the temperature feedback alone. The results shown in Fig.~\ref{fig:fig6}E indicate that eigenmodes with large eigenvalues, corresponding to fast-decaying modes, exhibit large perturbations in the lower troposphere and those with small eigenvalues have large perturbations in the upper atmosphere.

All temperature-feedback circuit gains (Fig.~\ref{fig:fig6}D) satisfy $0<g_i<1$. The positivity of the temperature-feedback circuit gains indicates that temperature feedback acts to amplify imposed energy perturbations, while their amplitudes being less than unity ensures that a fraction of the unbalanced energy perturbation entering each temperature-feedback circuit can always be radiated to space as outgoing longwave radiation (OLR) during each cycle. As a result, the unbalanced energy perturbations returned to each individual layers are progressively reduced and the system converges monotonically toward equilibrium at which the imposed energy perturbations are amplified at all layers. The work of ~\cite{cai2024, hu2025, sun2025} derived the equilibrium solution of temperature response to imposed energy perturbations from the energy gain kernel (EGK) associated with the Planck feedback matrix. We have confirmed that the temporal evolution of energy perturbations derived from the temperature feedback-circuit decomposition exactly converges to EGK. Therefore, for the atmosphere-surface climate system represented by the Planck feedback matrix, we refer to the corresponding forcing-transformation kernels for a general coupled linear system as the first-cycle energy gain kernel and energy gain kernel, respectively.

The condition $0<g_i<1$ also implies that all pre-cycle pathways in the temperature-feedback circuits are direct, so that the first-cycle energy kernel is identical to the matrix $\U$ (Fig.~\ref{fig:fig6}F). Because $0<g_i<1$, amplification of the first-cycle energy gain kernel through successive feedback cycles is governed by the loop gains, which are identical to the circuit gains. This leads to monotonic convergence with increasing energy perturbation magnitudes toward the energy gain kernel, with the values in each row equal to those of the corresponding row in the first-cycle energy gain kernel multiplied by the amplification factor $1/(1-g_i)>1$ specific to that row.  As a result, the smooth positive-definite structures in the first-cycle energy gain kernel, which naturally emerge from radiative coupling within the coupled atmosphere–surface column, are proportionally amplified to the energy gain kernel. Again, this demonstrates that, for the atmosphere-surface climate system, the rows of the first-cycle energy gain kernel serve a role analogous to eigenvectors, whereas the temperature-feedback circuit gains serve a role analogous to eigenvalues by determining the convergence rate and final amplification of the initial energy input to the atmosphere-surface column at equilibrium. 

\section*{Discussions}

This study identifies a complete set of independent intrinsic feedback circuits, one for each system component, in an invertible general linear system that has nonzero diagonal entries and nonsingular principal submatrices. An individual feedback circuit serves as a source-centered circuit when externally imposed forcing is applied at the corresponding system component and as a non-source-centered circuit when externally imposed forcing is applied at other system components.  Each feedback circuit consists of two integral components: an intrinsic circuit gain and a pair of forcing-transfer factors. Intrinsic circuit gains are determined from within-circuit forcing transfers. Associated with each intrinsic circuit gain are the loop gain and pre-cycle forcing-transfer factor, which are introduced to ensure the convergence of successive feedback cycles while preserving the magnitude of the within-circuit forcing transfer for both regular and over-compensating circuits. The other forcing-transfer factor characterizes the cross-circuit forcing transfer from a source-centered circuit to its corresponding non-source-centered circuits. All intrinsic circuit gains and cross-circuit forcing-transfer factors are pre-determined based on unit forcing imposed at individual system components and therefore represent intrinsic circuit properties of the linear coupled system under consideration. Because these intrinsic circuit properties are not forcing specific, they can therefore be directly applied to externally imposed forcing with arbitrary finite amplitudes acting on the same coupled system without further scaling.

The pair of pre-cycle and cross-circuit forcing-transfer factors for each system component forms the corresponding elements of a pair of forcing-transfer matrices, pre-cycle and cross-circuit forcing-transfer matrices. The product of these two forcing-transfer matrices is the first-cycle forcing-transformation kernel, which provides the analytical representation of the transformed forcing at the first feedback cycle. Each row of the first-cycle forcing-transformation kernel serves a role, in parallel with an eigenvector, for linear forced problems in characterizing the structure of the transformed forcing at the component corresponding to the row due to both within-circuit and cross-circuit forcing transfers from all individual components. The corresponding loop gain serves a role, in parallel with the associated eigenvalue, characterizing whether the temporal evolution of the transformed forcing at that component, through successive feedback cycles, exhibits a monotonic convergence with increasing magnitude or an oscillatory convergence with decreasing magnitude toward the equilibrium transformed forcing. Because all loop gains, which are defined from their corresponding intrinsic circuit gains, have magnitudes less than unity, the first-cycle forcing-transformation kernel converges exactly to the equilibrium forcing-transformation kernel through successive cycles of individual feedback circuits, regardless of whether the system contains unstable eigenmodes. 

Under the feedback-circuit decomposition, the forcing balance at individual components is accounted for separately for the original externally imposed forcing and its transformation. The original externally imposed forcing is balanced at other components within the source-centered circuit, as manifested by simultaneous successive reduction in both the unbalanced forcing returned to the corresponding source component and the unreturned forcing that constitutes the forcing output outside the system. Such successive reduction follows the geometric series associated with the loop gain of the corresponding source circuit, which vanishes at equilibrium so that the total forcing output is exactly equal to the original externally imposed forcing. Because the balance of the external forcing is in the form of forcing output, it cannot be expressed in terms of the system response. The transformed forcing at each individual component is balanced locally within its own circuit. Therefore, the temporal solution for the system response naturally emerges from transformed forcing at individual components normalized by diagonal elements of the system matrix. The convergence of the first-cycle forcing-transformation kernel to the equilibrium kernel automatically ensures the convergence of the transient response to the equilibrium response. In light of the above, the feedback-circuit decomposition is therefore intrinsically conservation-based and naturally leads to the causally driven temporal evolution of the system response to externally imposed forcing.

The validity of the conservation-based feedback-circuit decomposition is demonstrated through two representative, but randomly generated general coupled systems, one containing unstable eigenmodes and the other without unstable eigenmodes. Each of the two cases is used to illustrate the characteristics of positive and negative regular circuits, and positive and negative over-compensating circuits. The demonstrations show that the feedback-circuit decomposition provides exact analytical representations of transformed forcing and exact local balance at every feedback cycle while converging exactly toward equilibrium transformed forcing regardless of the existence of unstable eigenmodes. For the stable case, both analytical solutions derived from the feedback-circuit decomposition and the corresponding time-integration solutions converge to the same equilibrium state. 

The conservation-based feedback-circuit decomposition is further illustrated through a real-world application to the Planck feedback matrix. The Planck feedback matrix represents the coupled linear system for temperature feedback in the coupled atmosphere–surface climate system, in which the abstract forcing-transformation kernel becomes the energy gain kernel once the underlying physics is specified. In a separate study, we further demonstrate that climate sensitivity emerging from the temperature-feedback-circuit decomposition can accurately predict global-mean warming responses to imposed radiative forcing at the top of the atmosphere without requiring conventional forward integration of climate models.

In essence, the conservation-based feedback-circuit decomposition untangles the highly intertwined coupling in the original linear system and isolates analytically the underlying distinct forcing transfer pathways in their own specific forms among different system components, allowing transformed forcing to be obtained analytically while preserving forcing balance throughout successive feedback cycles. The conservation-based feedback-circuit decomposition solves linear forced problems by explicitly predicting how the initially imposed forcing evolves in the phase space of coupled system components, rather than through temporal evolution in physical time. Because the evolving forcing is balanced instantaneously and exactly within individual feedback circuits, the imposed forcing and its transformation can continuously evolve through successive feedback cycles until the geometric series associated with the loop gains converges, at which point the feedback circuits cease to operate and equilibrium is established. Therefore, as long as the imposed forcing is finite and the mild algebraic conditions required for the existence of a complete set of feedback circuits are satisfied, the forcing transformation can always eventually terminate at the equilibrium transformed forcing, regardless of whether the corresponding dynamical system is stable or unstable with respect to random perturbations or to the projections of the imposed forcing onto unstable eigenmodes. In this sense, the conservation-based feedback-circuit decomposition provides a complementary representation of coupled linear systems from the eigenmode decomposition. The complementary aspect is particularly evident for systems prone  to instability for which the feedback-circuit decomposition is capable of revealing the convergence toward the equilibrium response exactly without identifying instability. The eigenmode decomposition, on the other hand, reveals the instability but without identifying the convergence to equilibrium. 

The equilibrium forcing-transformation kernel can be directly obtained from the inverse of the linear system matrix. The relationships between the inverse matrix and the feedback-circuit quantities derived from the feedback-circuit decomposition can be numerically recovered directly from the full system matrix, without explicitly inverting reduced subsystems. These quantities include the cross-component forcing transfer matrix, intrinsic circuit gains and their associated loop gains and pre-cycle forcing transfer matrix, and the first-cycle forcing-transformation kernel. This provides a practical and computationally efficient pathway for applying the feedback-circuit decomposition to high-dimensional coupled systems without requiring eigenmode or full spectral decomposition.

The conservation-based feedback-circuit decomposition provides a new generic mathematical tool and an alternative physically interpretable framework for theoretical advancement and practical applications across coupled scientific and engineering systems. By explicitly identifying the intrinsic forcing-transfer and gain structures embedded within coupled systems, the framework may offer new perspectives for understanding, predicting, and controlling forced responses in areas involving complex interactions and feedback processes. More broadly, the physically interpretable transfer and balance structures revealed by the feedback-circuit decomposition may provide useful insights for the design and interpretation of external feedback and control processes in scientific and engineering applications.

\clearpage

\begin{figure}
\centering
\includegraphics[width=0.78\textwidth]{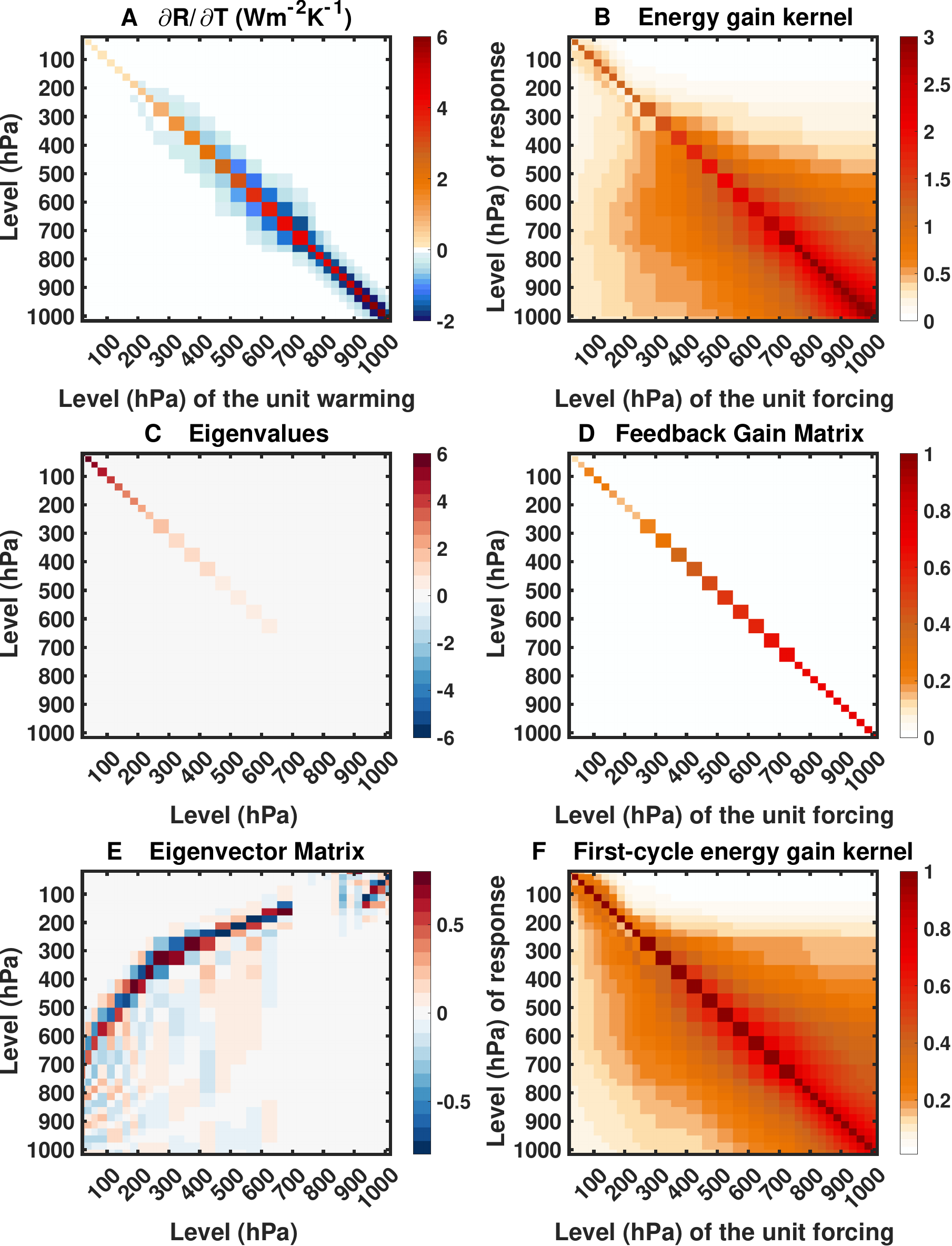}
\caption{\textbf{Example 3 of feedback-circuit decomposition for the Planck feedback matrix $\partial \mathbf{R}/\partial \mathbf{T}$.}
(\textbf{A}) Planck feedback matrix $\partial R/\partial T$. (\textbf{B}) Energy gain kernel. (\textbf{C}) Eigenvalue matrix. (\textbf{D}) Feedback gain matrix. (\textbf{E}) Eigenvector matrix. (\textbf{F}) First-cycle energy gain kernel. It is noted here that, for convenience, pressure levels are used in both coordinates of panel (\textbf{C}), as well as in the horizontal coordinate of panel (\textbf{E}), only for index ordering and bear no physical meaning in these specific coordinates of the two panels.}
\label{fig:fig6}
\end{figure}

\clearpage

\section*{Acknowledgments}
The author thanks Dr. Jie Sun for numerous helpful discussions and constructive comments on an early version of the paper.

\paragraph*{Funding:}
The author received no specific funding for this work.

\paragraph*{Author contributions:}
M.C. conceived the study, developed the theory, performed the analysis, and wrote the manuscript.

\paragraph*{Competing interests:}
The author declares no competing interests.

\paragraph*{Data and materials availability:}
The ERA5 reanalysis is available at \url{https://doi.org/10.5065/P8GT-0R61}. Model codes and data analysis codes will be made available upon reasonable request to enable reproduction of the results.

\end{document}


\begin{center}
{\large \textbf{Supplementary Materials for}}\\[0.5em]
{\Large \textbf{Conservation-Based Feedback-Circuit Decomposition for Linear Forced Systems}}\\[1em]
Ming Cai\\
Department of Earth, Ocean, and Atmospheric Science\\
Florida State University\\
Tallahassee, Florida 32306\\
Email: \href{mailto:mcai@fsu.edu}{mcai@fsu.edu}
\end{center}

\section*{A. Data}

Monthly mean data from the European Centre for Medium-Range Weather Forecasts (ERA5) reanalysis~\cite{Hersbach2020} covering the period of 1980--2000 are used to construct climatological mean vertical profiles of zonal-mean atmospheric temperature, specific humidity, clouds, and ozone, as well as zonal-mean surface albedo, surface pressure, and surface temperature. The ERA5 reanalysis is available from the Climate Data Store (\url{https://cds.climate.copernicus.eu/}). The horizontal resolution of the dataset is $1.5^{\circ}\times1.5^{\circ}$ in longitude and latitude, with 37 pressure levels plus the surface level. Only data above the local surface pressure are retained in all calculations.

\section*{B. Construction of the Planck feedback matrix}

We apply the radiative transfer model~\cite{FuLiou1992,FuLiou1993} to calculate the vertical profile of longwave (LW) cooling rates, in units of W~m$^{-2}$, using input fields derived from the ERA5 1980--2000 mean state, denoted as $\mathbf{R}^{(1980-2000)}$. Following~\cite{LuCai2009,CaiLu2009}, we obtain the Planck feedback matrix, $\partial R_i/\partial T_j$, in units of W~m$^{-2}$~K$^{-1}$, by repeating the radiative transfer model calculations $L$ times, where $L$ denotes the total number of layers in a given atmosphere--surface column, with layer $L$ representing the surface layer and layer 1 the top layer. Each of the $L$ radiative transfer model calculations is made with a 1~K warming perturbation applied to the $j$-th layer and all other variables held fixed at their 1980--2000 climatological mean values. The resulting vertical profile of the LW cooling rates is denoted by $\mathbf{R}^{(j)}$. The difference between $\mathbf{R}^{(j)}$ and $\mathbf{R}^{(1980-2000)}$ is the vertical profile of the radiative cooling rates due to a 1~K warming at the $j$-th layer, corresponding to the $j$-th column of $\partial R_i/\partial T_j$, namely,
\begin{equation}
\frac{\partial R_i}{\partial T_j}=R_i^{(j)}-R_i^{(1980-2000)},\quad \text{for } i=1,2,\ldots,L .
\end{equation}
Repeating this procedure for all layers yields the full matrix of $\partial R_i/\partial T_j$.

\section*{C. Supplementary Figures}

\begin{figure}[p]
\centering
\includegraphics[width=\textwidth]{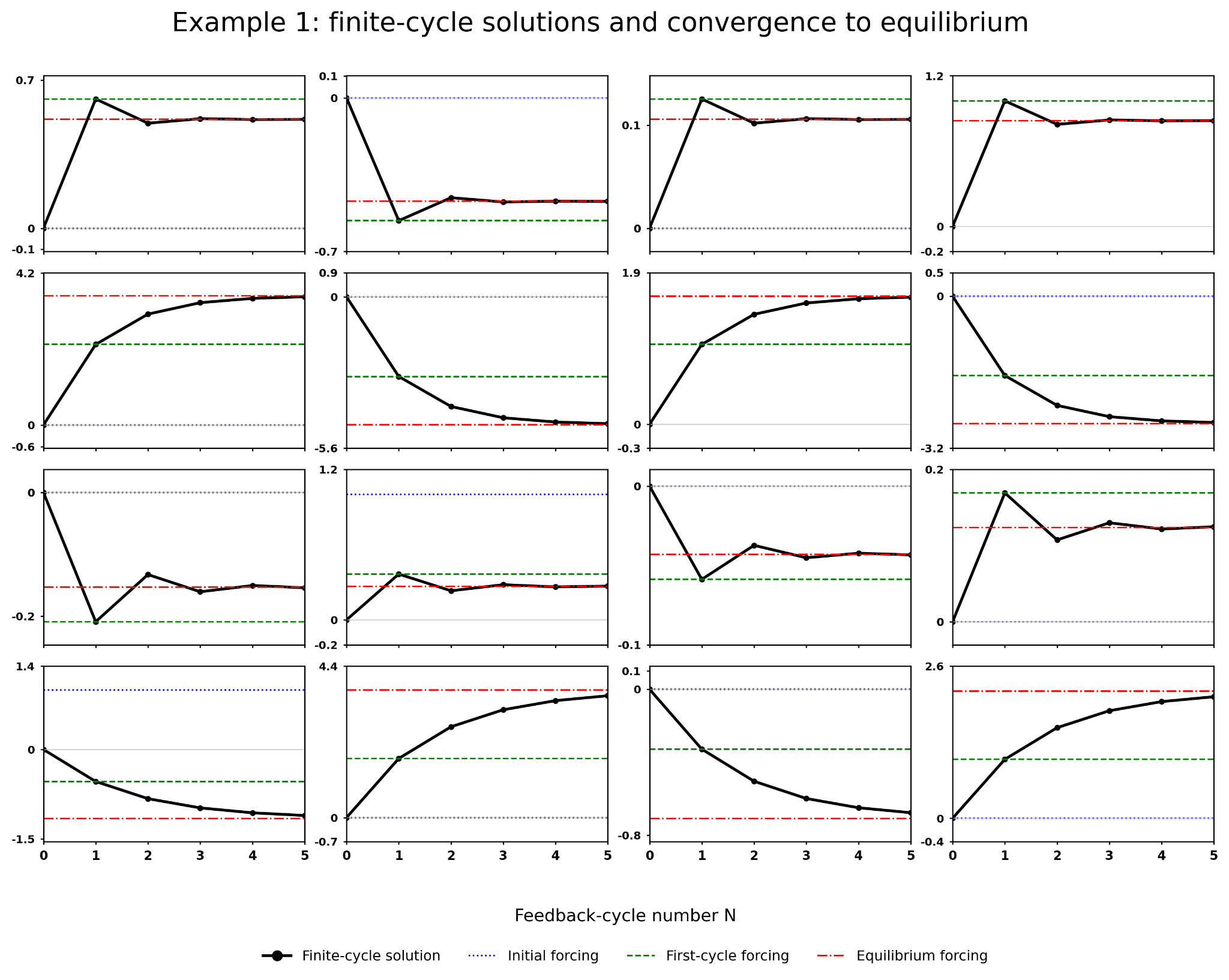}
\caption{\textbf{Finite-cycle solutions and convergence to equilibrium for Example 1 (stable $4\times4$ coupled linear system).}
Columns represent system components 1--4 from left to right, with the initial input forcing indicated by blue dotted lines. Rows represent the finite-cycle analytical solutions of the amplified forcing (black curves, given by $\bm{\Lambda}^{(N)}\mathbf{F}$) for system components 1--4 from bottom to top in response to the initial forcing imposed at the corresponding column components. The green dashed and red dash-dotted lines denote the first-cycle and equilibrium solutions for the amplified forcing, $\bm{\Lambda}^{(1)}\mathbf{F}$ and $\bm{\Lambda}\mathbf{F}$, respectively. The abscissa represents feedback-cycle number, and the ordinate represents forcing values.}
\label{fig:S1}
\end{figure}

\begin{figure}[p]
\centering
\includegraphics[width=\textwidth]{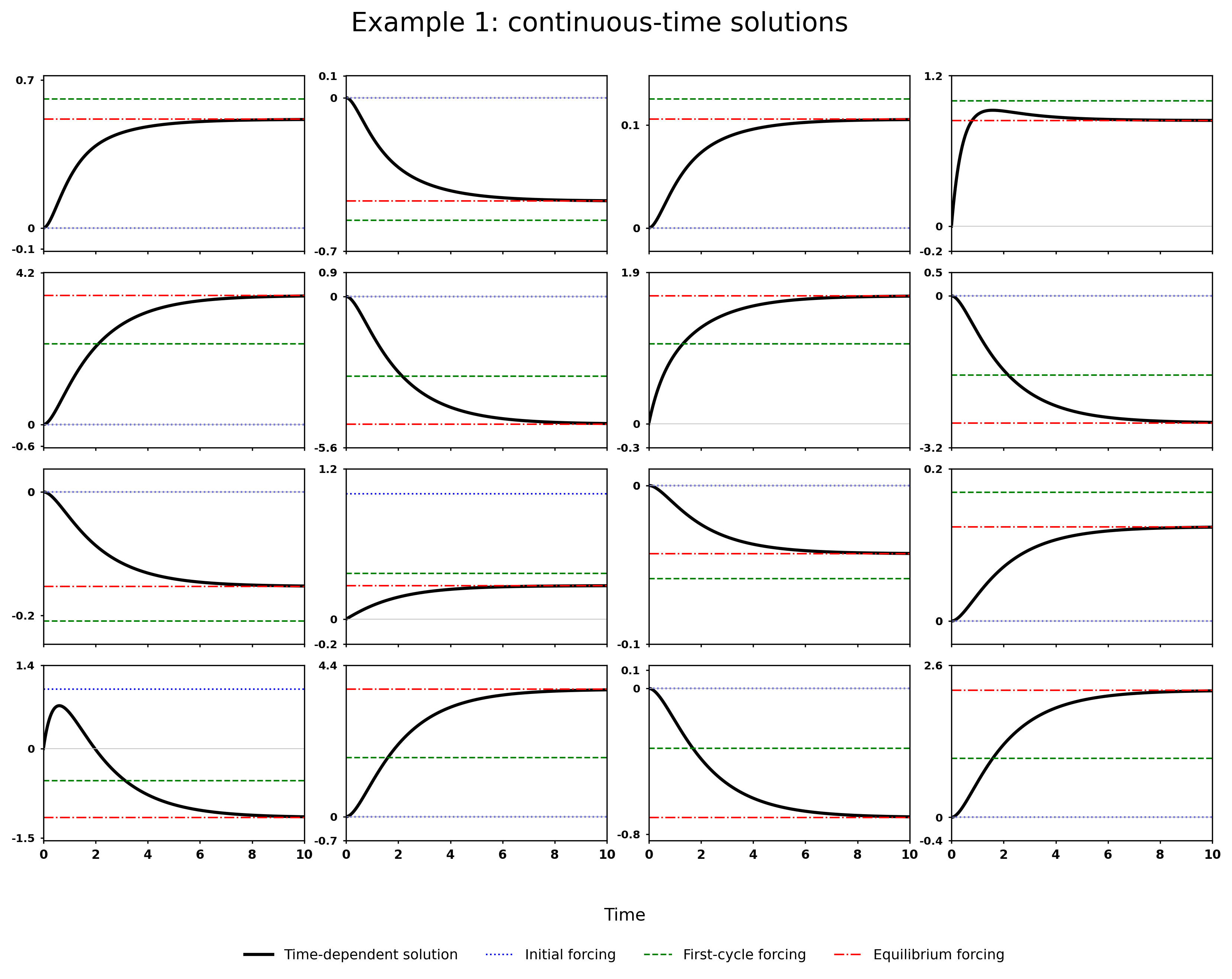}
\caption{\textbf{Continuous-time solutions and convergence to equilibrium for Example 1 (stable $4\times4$ coupled linear system).}
All information is identical to that in Fig.~\ref{fig:S1}, except the time-dependent solutions are obtained numerically with the abscissa representing time. Because the numerical solutions do not directly yield the amplified forcing, we deliberately plot $\operatorname{diag}(\mathbf{A})\mathbf{X}(t)$, instead of $\mathbf{X}(t)$, for a direct comparison with the analytical solutions shown in Fig.~\ref{fig:S1}.}
\label{fig:S2}
\end{figure}

\begin{figure}[p]
\centering
\includegraphics[width=\textwidth]{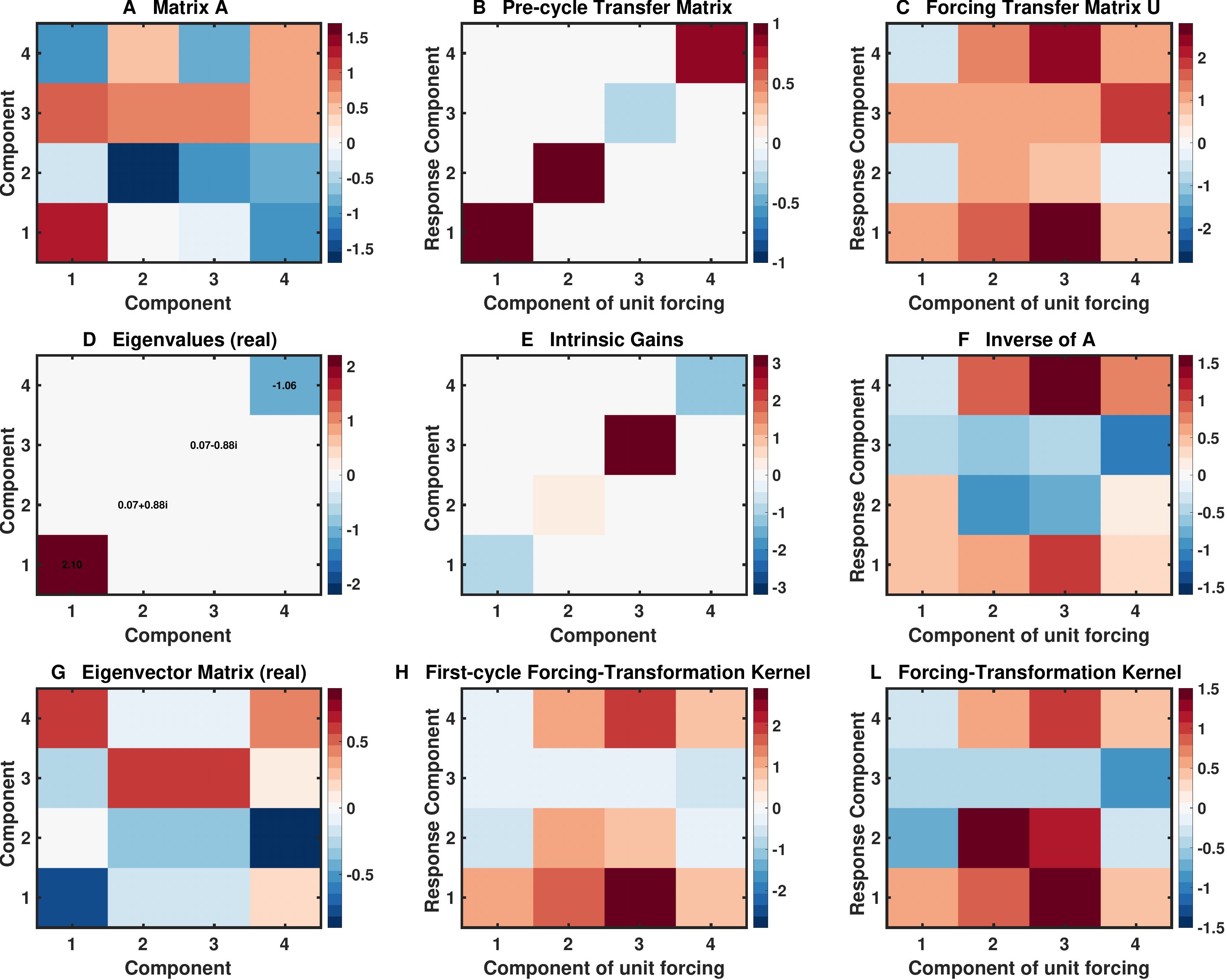}
\caption{\textbf{Unstable example of feedback-circuit decomposition for a randomly generated unstable $4\times4$ linear system.}
Same as Fig.~3 except for the unstable $4\times4$ linear system whose system matrix is given in panel A. In panel D, the numbers along the diagonal direction are the eigenvalues of the system. As in Fig.~3, eigenvalues are displayed using the same component ordering as the eigenvectors and circuit quantities for convenience, but this ordering in both coordinates in panel D, as well as in the horizontal coordinate of panel G, bears no physical meaning.}
\label{fig:S3}
\end{figure}